\documentclass[12pt]{article}
\usepackage{jhep-mod-no-toc}
\usepackage{bm}
\usepackage{amssymb,amsmath,amsthm}
\usepackage{mathrsfs}
\usepackage[utf8]{inputenc}
\usepackage[pdftex]{hyperref}
\hypersetup{colorlinks=true} 
\usepackage{xcolor}
\usepackage{appendix}
\usepackage{graphicx}
\usepackage{dcolumn}
\usepackage{bm}
\usepackage{multirow}
\usepackage{float}
\usepackage[normalem]{ulem}
\usepackage{graphicx}
\usepackage[T1]{fontenc}
\def\setskip{}
\begin{document}
\newcommand{\perhMpcUnit}{$h^{-1}$\,Mpc}
\newcommand{\perhMpc}[1]{#1\,\perhMpcUnit{}}
\newcommand{\reff}{R_\mathrm{eff}}
\newcommand{\gevolution}{\textit{gevolution}}
\newcommand{\LCDM}{$\Lambda$CDM}
\newcommand{\threericci}{\tensor[]{\mathcal{R}}{}}
\newcommand{\Hall}{\bar{H}}
\newcommand{\Hm}{\bar{H}_0}
\newcommand{\OmegaCurvature}{\Omega_\mathrm{K}}
\newcommand{\OK}{\bar\Omega_\mathrm{K}}
\newcommand{\OM}{\bar\Omega_\mathrm{M}}
\newcommand{\OR}{\bar\Omega_\mathrm{R}}
\newcommand{\OQ}{\bar\Omega_\mathcal{Q}}

\def\th{\theta}\def\ph{\phi}\def\rh{\rho}\def\si{\sigma}
\def\w#1{\,\hbox{#1}}
\def\kmsMpc{\w{km}\;\w{sec}^{-1}\w{Mpc}^{-1}}
\definecolor{purple}{rgb}{1,0,1}
\newcommand{\red}[1]{{\slshape\color{red} #1}}
\newcommand{\blue}[1]{{\slshape\color{blue} #1}}
\newcommand{\purple}[1]{{\slshape\color{purple} #1}}
\parindent0pt
\parskip7pt
\allowdisplaybreaks
\title{\Large{Solution to the cosmological constant problem}}
\author{\large David L. Wiltshire$^*$}
\affiliation{School of Physical \& Chemical Sciences, University of Canterbury,\\ Private Bag 4800, Christchurch 8140, New Zealand}
\emailAdd{David.Wiltshire@canterbury.ac.nz}
\abstract{\setskip \\
I propose an observationally and theoretically consistent resolution of the cosmological constant problem: $\Lambda$ is a counterterm -- with a running coupling -- that balances the monopole celestial sky average of the kinetic energy of expansion of inhomogeneously distributed `small' cosmic voids. No other dark energy source is required. This solution relies on the first investigation of void statistics in cosmological simulations in full general relativity (arXiv:2403.15134). Results are consistent with parameters of the Timescape model of cosmological backreaction. Crucially, dynamical spatial curvature arises as time-varying spatial gradients of the kinetic spatial curvature, and depends directly on the void volume fraction. Its monopole average generates the cosmological term. This result potentially resolves the Hubble tension and offers new approaches to tackling other tensions and anomalies.

\medskip\noindent
$^*$ OrcidID: 0000-0003-1992-6682
		
\medskip\noindent
{\sc Date:} 31 March 2024; LaTeX-ed \today.

\medskip\noindent
\vspace{0.5cm}

{\emph{Essay written for the Gravity Research Foundation 2024 Awards\\ for Essays on Gravitation.}}

}
\notoc
\maketitle
\clearpage \setskip

The road to a theory of quantum gravity has twisted and turned for decades, through a cornucopia of fascinating ideas. Such ideas have kick-started whole fields of mathematics, precision experiments and space-based technologies. But science is data-driven. To begin to unveil quantum gravity we have had to wait until the century of multi--messenger astronomy, for the data needed to formulate the right questions about the Universe, its origin and evolution.


While fundamental approaches including string theory and loop quantum gravity have taught us a lot, none has presented a compelling solution to the question of the vacuum ground state of the Universe. Rather, decades of thought have left us not much further than Weinberg's anthropic answer to the cosmological constant problem \cite{Weinberg89}. In this essay I propose a new solution, one that may shortly be tested with 21st century data. The solution involves a model of cosmological backreaction, the Timescape \cite{clocks,sol,obs}, developed in the 2000s. The key new ingredients are the quantitative results of the first investigation of void statistics in cosmological simulations with full numerical general relativity \cite{Williams24}. These required development of new analysis tools, adapting those used both for Newtonian $N$-body simulations and data analysis.\footnote{\setskip M Williams has developed a new watershed void finder \cite{Williams24}, combining methodologies of different analysis tools and the empirically-derived
Hamaus--Sutter--Wandelt (HSW) density profiles \cite{HSW}. }

A crucial observation concerning the late epoch Universe is that on scales less than at least 150--300 Mpc it displays a complex cosmic web of structure, dominated in volume by voids. Voids are threaded by expanding filaments containing galaxy clusters and surrounded by denser sheets (walls). The densest locations are knot-like intersections tens of Megapasecs across. Understanding the quasilocal gravitational energy gradients of such complex environments is central to the Timescape cosmology. These issues have been discussed in past essays \cite{GRF07,FQxI08}. The present discussion concerns the primary focus in both the Timescape and our numerical relativity (NR) simulations: the largest \emph{typical} voids, which I heuristically call \emph{small voids}.\footnote{\setskip Some 40\%
of the volume of the present universe is in voids of a characteristic diameter \perhMpc{30} \cite{HV,Pan11}, where $h$ is the dimensionless Hubble constant, $\Hm=100h\kmsMpc$. Adding numerous smaller minivoids and rare larger voids increases the average present void volume fraction, $f_{{\rm v}0}$.} NR simulations may well provide a rigorous framework for quantifying just how atypical larger structures are. In fact, new observations of large structures are increasingly at odds with the standard Friedmann--Lema\^{\i}tre--Robertson--Walker (FLRW) cosmology \cite{CPrev22},\footnote{\setskip It should be stressed that atypical structures, including the newly discovered ``Big Ring'' \cite{LCW24} are \emph{expanding structures} in the category of filaments / sheets, each containing galaxy clusters -- the largest gravitationally bound structures. Their Timescape interpretation is fundamentally different to \LCDM.} and its $\Lambda$ Cold Dark Matter (\LCDM) variant. While such observations may call into question naming any \LCDM\ model a ``concordance cosmology'' they are not the primary concern of this essay. Rather, in setting aside a FLRW framework long past its use-by date, a better framework first needs to be sketched. 

It is conventionally assumed that departures from average isotropic cosmic expansion always reduce exactly to local Lorentz boosts -- i.e., peculiar velocities of the source, observer and intervening structures -- on a global FLRW background. However, in GR the growth of structure may lead to an expansion history significantly different to any single global FLRW background, giving rise to \emph{backreaction of inhomogenities}. Such cosmological backreaction scenarios 
may in turn affect the traditional kinematic interpretation of peculiar velocities \cite{bnw16}.

Different types of backreaction have been classified in the well-known Buchert formalism \cite{Buchert20,BMR},
which provides a set of equations for the evolution of scalar averages of small scale structures.
In the Timescape model, with Buchert averages applied to the entire particle horizon volume, ${\mathcal V}_{\mathcal H}$, the first Buchert equation for average cosmic evolution may be written as the
{\em energy density sum-rule}:
\begin{equation}
\label{eq:sumrule}
\OM+\OR+\OK+\OQ=1\,.
\end{equation}
Here $\OM$ and $\OR$ are analogous to matter and radiation energy densities in the Friedmann equation, but scale with respect to powers of an {\em average volume scale} $\bar a$, 
{\em not} a background metric scale factor; $\OK$ is the average spatial curvature or {\em kinetic curvature} density; and $\OQ$ is the {\it kinematical backreaction} density\footnote{\setskip Additional backreaction terms are also found in general \cite{BMR}, including a cosmological constant backreaction density, $\bar\Omega_\Lambda$. The latter is irrelevant for the hypothesis of this essay.
At late epochs the radiation density $\OR$ is only relevant insofar as the ratio $\OR/\OM\simeq0.3$ at last scattering, with significance for calibrations relative to CMB data.
This notation differs from ref.\ \cite{BMR}, where $\Omega_\mathcal{R}$ is used in place of $\OmegaCurvature(z,\vartheta,\varphi)$ and no radiation density is explicitly written down.
Positive values $\OK=\langle\OmegaCurvature\rangle_{{\mathcal V}_{\mathcal H}}>0$ correspond to negative spatial curvature, as is the case for void domination.
}.

Since the 2000s much attention has focused on the magnitude of 
$|\OQ|$. The Timescape model was designed to address criticisms of Ishibashi and Wald \cite{IW06}, showing that while nonzero kinematical backreaction
is necessary for realistic cosmology constrained by the initial matter power spectrum, $|\OQ|$ need not be large \cite{clocks,sol,obs}. With radiation included, Planck CMB constraints are consistent with a kinematical backreaction magnitude of $|\OQ|\sim10^{-6}$--$10^{-5}$ at last scattering that evolves to a maximum $|\OQ|\sim0.03$--$0.04$ \cite{dnw}.

\begin{figure}[!h]
	\begin{center}
		\includegraphics[scale=0.44]{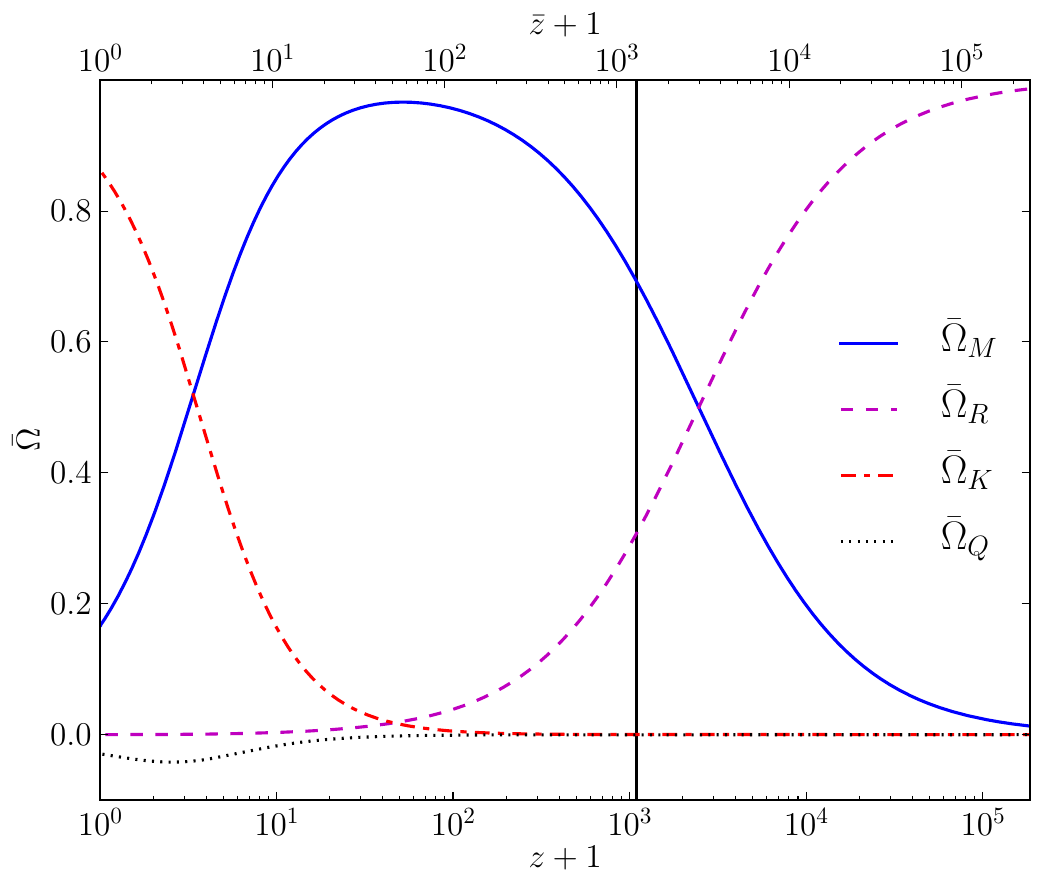}
		\vspace{-0.5cm}
		\caption{\label{fig:sumrule} Energy density parameters in eq.~(\ref{eq:sumrule}) for Timescape, fit to
{\em Planck} data, as a function of redshift (from ref.\ \cite{dnw}, Fig.~1). The vertical line indicates the CMB surface (decoupling). The parameter $\bar z$ (top horizontal scale) is the bare redshift, while the parameter $z$ (lower horizontal scale) is the dressed redshift that coincides with our observations.}
	\end{center}	
\end{figure}

In the Timescape model, the void volume fraction, $f_{\rm v}$, is the only free function and $\OM$ is also related to it \cite{obs}. Furthermore, $\OK\propto f_{\rm v}^{1/3}/(\bar a^2\bar H^2)$ has a direct physical interpretation as the kinetic energy of expansion of voids\footnote{\setskip It is a generic feature of most realistic attempts to model backreaction, that rigidly varying FLRW spatial curvature $\Omega_{\rm K\,FLRW}=-k/(\bar a^2\bar H^2)$, $k=\,$const, is replaced with significant emerging negative spatial curvature terms, even if not completely replacing $\Lambda$ \cite{B18}.} (see Fig.~\ref{fig:sumrule}). 

The new void statistics methodology of ref.\ \cite{Williams24} has enabled the first direct identification of voids in NR simulations. The same simulations were analysed in the past without explicit determination of the spatial curvature within voids \cite{MLP18,MLP19}. In particular, new results can be expected -- and have been found \cite{Williams24} -- by careful identification of voids, as opposed to selecting random volumes of both voids and filaments. A rigorous numerical understanding of the asympototic approach to an isotropic average expansion is a key criterion for understanding the numerical procedures of stacking voids. The procedures adopted in ref.\ \cite{Williams24} were empirically determined, and adapted established methodologies \cite{HSW}. The procedures did not incorporate any particular cosmology, but allowed for a generic non--FLRW average expansion as well as FLRW average expansion.

\begin{figure}[!h]
	\begin{center}
		\includegraphics[scale=0.99]{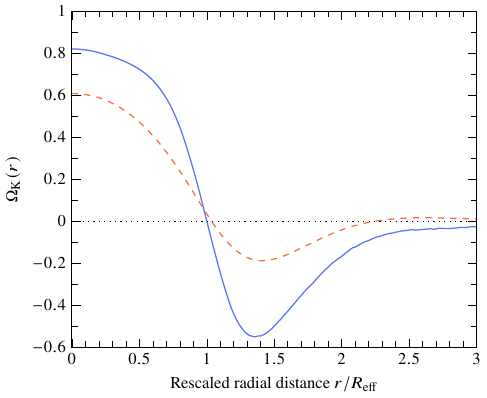}
		\vspace{-0.5cm}
		\caption{\label{fig:radial_omega} $\OmegaCurvature(r)$ 
                  averaged over shells of radius $r$ centred on stacked void centres, each void scaled by its effective radius $\reff$.
    The solid (dashed) curve shows results of a \perhMpc{4} (\perhMpc{12}) resolution simulation. (See ref.\ \cite{Williams24}, Fig.~9.)}
	\end{center}	
\end{figure}

The results are striking since the signature of large negative spatial curvature, common to several backreaction scenarios with emerging spatial curvature, is evident in the radial profile of the stacked voids (see Fig.~\ref{fig:radial_omega}). Furthermore, for the finest \perhMpc{4} resolution available in the NR simulations the magnitude of the spatial curvature at the void centres is consistent numerically with Timescape model estimates based solely on the angular scales of the CMB acoustic peaks in the 2013 Planck data release, namely $\OK=0.862^{+0.024}_{-0.032}$, $\OM=0.167^{+0.036}_{-0.037}$, $\OQ=-0.0293^{+0.0033}_{-0.0036}$ at present, $z=0$ (see Fig.~\ref{fig:sumrule} and Table~1 of \cite{dnw}).

Ostensibly this is surprising, since the same simulations had initial conditions chosen to give an Einstein--de Sitter model with no dark energy if FLRW average evolution is self-consistent. Indeed, if one takes averages over randomly chosen regions that contain both voids and walls, the magnitude of the average spatial curvature appears to be small \cite{MLP19}. Further finer resolution studies deep within voids, unbound to any cosmic structures, could allow for particular tests of the Timescape model.

The ostensibly surprising results do, however, suggest an actual solution of the cosmological constant problem, rather than merely of the cosmic coincidence problem \cite{clocks,GRF07}. An alternative weak field $N$--body approach to large cosmological simulations, \gevolution, in which expansion is prescribed by scaling the simulation volume according to the Friedmann equation from an FLRW model has also been interpreted as disfavouring significant cosmological backreaction. However, \gevolution\ operates by defining a renormalized scale factor \emph{and a renormalized time coordinate} such that the new spatially average line element looks like an unperturbed Friedmann model \cite{ADDK}\footnote{See, in particular, ref.\ \cite{ADDK} Sec.\ 5.3.}.

If my hypothesis is correct, then the \gevolution\ renormalization to subtract a spatially average monopole -- in a procedure involving the time coordinate -- is the numerical signature of phenomenological methods familiar to the study of running couplings in quantum field theory. Backreaction and renormalization group methods extend beyond mere analogies of language,\footnote{The Timescape model also involves bare and dressed parameters -- the latter correspond to our own observations but differ from their FLRW counterparts and do not obey a simple sum rule (\ref{eq:sumrule}).} and are key to a deeper geometrical understanding of the nature of the vacuum of quantum gravity. The timescape phenomenology, which exploits a conformal ambiguity in the time parameter associated with light cone averages, is also key. At a time when new theoretical advances in understanding gravitational wave production invoke scattering amplitudes on one hand, and role of the Bondi--Metzner--Sachs group in defining null infinity on the other, it appears that we may truly be on the verge of substantial breakthroughs. However, it requires teamwork on an unprecedented scale. In particular, it requires putting to rest the 100--year FLRW assumption that never was a part of the foundations of Einstein's general relativity.

\begin{figure}[!h]
	\begin{center}
		\includegraphics[scale=0.67]{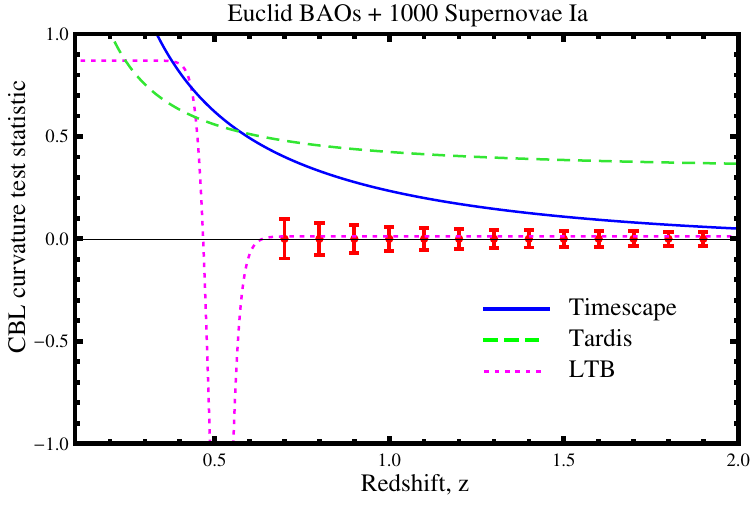}
		\vspace{-0.5cm}
		\caption{\label{fig:curveball} Test of FLRW curvature. Simulated \LCDM\ data \cite{smn} (red points) for the Euclid mission 
  with 2 falsifiable backreaction predictions -- Timescape \cite{obs}, Tardis \cite{tardis} -- and an unfalsifiable Lema\^{\i}tre--Tolman--Bondi (LTB) large void prediction \cite{gbh}.}
	\end{center}	
\end{figure}

The key ideas presented here are not intended merely to provoke. Over a decade ago, independent projections were made \cite{smn} for the Euclid mission (see Fig.~\ref{fig:curveball}) to directly test the Friedmann equation for average evolution via the Clarkson--Bassett--Lu (CBL) test \cite{CBL}. These projections also have the precision to distinguish an FLRW expansion history from the particular non-FLRW expansion histories of the Timescape and Tardis models. Now that Euclid has finally been successfully launched, we can expect definitive answers within a few years.

\section*{Acknowledgements}
I thank Michael Williams and Hayley Macpherson for helpful discussions. From April 2024 onwards, our work will be supported by the Marsden Fund, administered by the Royal Society Te Ap\=arangi of Aotearoa / New Zealand. 



\begin{thebibliography}{25}

\bibitem{Weinberg89}
S.~Weinberg,
``The cosmological constant problem,''
Rev.\ Mod.\ Phys.\ \textbf{61}, 1-23 (1989)

\bibitem{clocks}
D.~L.~Wiltshire,
``Cosmic clocks, cosmic variance and cosmic averages,''
New J.\ Phys.\ \textbf{9}, 377 (2007)
[arXiv:gr-qc/0702082]

\bibitem{sol}
D.~L.~Wiltshire,
``Exact solution to the averaging problem in cosmology,''
Phys.\ Rev.\ Lett.\ \textbf{99}, 251101 (2007)
[arXiv:0709.0732 [gr-qc]]

\bibitem{obs}
D.~L.~Wiltshire,
``Average observational quantities in the timescape cosmology,''
Phys.\ Rev.\ D \textbf{80}, 123512 (2009)
[arXiv:0909.0749 [astro-ph.CO]]

\bibitem{Williams24}
M.~J.~Williams, H.~J.~Macpherson, D.~L.~Wiltshire and C.~Stevens,
``First investigation of void statistics in numerical relativity simulations,''
arXiv:2403.15134 [astro-ph.CO]

\bibitem{HSW}
N.~Hamaus, P.~M.~Sutter and B.~D.~Wandelt,
``Universal density profile for cosmic voids,''
Phys.\ Rev.\ Lett.\ \textbf{112}, 251302 (2014)
[arXiv:1403.5499 [astro-ph.CO]]

\bibitem{GRF07}
D.~L.~Wiltshire,
``Gravitational energy and cosmic acceleration,''
Int.\ J.\ Mod.\ Phys.\ D \textbf{17}, 641-649 (2008)
[GRF Essay 2007; arXiv:0712.3982 [gr-qc]]

\bibitem{FQxI08}
D.~L.~Wiltshire,
``From time to timescape -- Einstein's unfinished revolution,''
Int.\ J.\ Mod.\ Phys.\ D \textbf{18}, 2121-2134 (2009)
[FQxI essay 2008; arXiv:0912.4563 [gr-qc]]

\bibitem{HV}
F.~Hoyle and M.~S.~Vogeley,
``Voids in the PCSZ survey and the updated Zwicky catalog,''
Astrophys.\ J.\ \textbf{566}, 641-651 (2002)
[arXiv:astro-ph/0109357];
``Voids in the 2dF Galaxy Redshift Survey,''
Astrophys.\ J.\ \textbf{607}, 751-764 (2004)
[arXiv:astro-ph/0312533]

\bibitem{Pan11}
D.~C.~Pan, M.~S.~Vogeley, F.~Hoyle, Y.~Y.~Choi and C.~Park,
``Cosmic voids in Sloan Digital Sky Survey Data Release 7,''
Mon.\ Not.\ Roy.\ Astron.\ Soc.\ \textbf{421}, 926-934 (2012)
[arXiv:1103.4156 [astro-ph.CO]]

\bibitem{CPrev22}
P.~K.~Aluri, 
\textit{et al.},
``Is the observable Universe consistent with the cosmological principle?,''
Class.\ Quant.\ Grav.\ \textbf{40}, 094001 (2023)
[arXiv:2207.05765 [astro-ph.CO]]

\bibitem{LCW24}
A.~M.~Lopez, R.~G.~Clowes and G.~M.~Williger,
``A big ring on the sky,''
arXiv:2402.07591 [astro-ph.CO]]

\bibitem{bnw16}
K.~Bolejko, M.~A.~Nazer and D.~L.~Wiltshire,
``Differential cosmic expansion and the Hubble flow anisotropy,''
JCAP {06}, 035 (2016)
[arXiv:1512.07364 [astro-ph.CO]]

\bibitem{Buchert20}
 T.~Buchert,
``On average properties of inhomogeneous fluids in general relativity I:
Dust cosmologies'',
Gen.\ Rel.\ Grav.\ \textbf{32}, 105-125 (2000)
[arXiv:gr-qc/9906015];
``On average properties of inhomogeneous fluids in general relativity II:
Perfect fluid cosmologies'',
Gen.\ Rel.\ Grav.\ \textbf{33}, 1381-1405 (2001)
[arXiv:gr-qc/0102049]

\bibitem{BMR}
T.~Buchert, P.~Mourier and X.~Roy,
``On average properties of inhomogeneous fluids in general relativity III: General fluid cosmologies,''
Gen.\ Rel.\ Grav.\ \textbf{52}, 27 (2020)
[arXiv:1912.04213 [gr-qc]]

\bibitem{IW06}
A.~Ishibashi and R.~M.~Wald,
``Can the acceleration of our universe be explained by the effects of inhomogeneities?,''
Class.\ Quant.\ Grav.\ \textbf{23}, 235-250 (2006)
[arXiv:gr-qc/0509108]

\bibitem{dnw}
J.~A.~G.~Duley, M.~A.~Nazer and D.~L.~Wiltshire,
``Timescape cosmology with radiation fluid,''
Class.\ Quant.\ Grav.\ \textbf{30}, 175006 (2013)
[arXiv:1306.3208 [astro-ph.CO]]

\bibitem{B18}
K.~Bolejko,
``Emerging spatial curvature can resolve the tension between high-redshift CMB and low-redshift distance ladder measurements of the Hubble constant,''
Phys.\ Rev.\ D \textbf{97}, 103529 (2018)
[arXiv:1712.02967 [astro-ph.CO]]

\bibitem{MLP18}
H.~J.~Macpherson, P.~D.~Lasky and D.~J.~Price,
``The trouble with Hubble: Local versus global expansion rates in inhomogeneous cosmological simulations with numerical relativity,''
Astrophys.\ J.\ Lett.\ \textbf{865}, L4 (2018)
[arXiv:1807.01714 [astro-ph.CO]]
\bibitem{MLP19}
H.~J.~Macpherson, P.~D.~Lasky and D.~J.~Price,
``Einstein\textquoteright{}s Universe: Cosmological structure formation in numerical relativity,''
Phys.\ Rev.\ D \textbf{99}, 063522 (2019)
[arXiv:1807.01711 [astro-ph.CO]]
  
\bibitem{ADDK}
J.~Adamek, D.~Daverio, R.~Durrer and M.~Kunz,
``gevolution: a cosmological N-body code based on General Relativity,''
JCAP {07}, 053 (2016)
[arXiv:1604.06065 [astro-ph.CO]]

\bibitem{smn}
D.~Sapone, E.~Majerotto and S.~Nesseris,
Phys.\ Rev.\ D \textbf{90}, 023012 (2014)
[arXiv:1402.2236 [astro-ph.CO]]

\bibitem{CBL}
C.~Clarkson, B.~Bassett and T.~H.~C.~Lu,
``A general test of the Copernican Principle,''
Phys.\ Rev.\ Lett.\ \textbf{101}, 011301 (2008)
[arXiv:0712.3457 [astro-ph]]

\bibitem{tardis}
M.~Lavinto, S.~R\"as\"anen and S.~J.~Szybka,
``Average expansion rate and light propagation in a cosmological Tardis spacetime,''
JCAP 12, 051 (2013)
[arXiv:1308.6731 [astro-ph.CO]]

\bibitem{gbh}
J.~Garcia-Bellido and T.~Haugboelle,
``The radial BAO scale and cosmic shear, a new observable for inhomogeneous cosmologies,''
JCAP {09}, 028 (2009)
[arXiv:0810.4939 [astro-ph]]

\end{thebibliography}
\end{document}